\documentclass[showpacs,superscriptaddress,aps,prl,twocolumn,notitlepage,floatfix,preprintnumbers,amsmath,amssymb]{revtex4-1}\usepackage{graphicx}
\usepackage{dcolumn}
\usepackage{subfigure}
\usepackage{bm}

 \usepackage{amsmath} 
\usepackage{comment}
\newcommand{\bmath}{\begin{displaymath}}
\newcommand{\emath}{\end{displaymath}}

\def\XXint#1#2#3{{\setbox0=\hbox{$#1{#2#3}{\int}$}
     \vcenter{\hbox{$#2#3$}}\kern-.5\wd0}}

\newcommand{\be}{\begin{equation}}
\newcommand{\ee}{\end{equation}}
\newcommand{\bea}{\begin{eqnarray}}
\newcommand{\eea}{\end{eqnarray}}

\newcommand{\p}{\partial}
\newcommand{\ba}{\begin{align}}
\newcommand{\ea}{\end{align}}

\usepackage{color}

\def\XXint#1#2#3{{\setbox0=\hbox{$#1{#2#3}{\int}$}
     \vcenter{\hbox{$#2#3$}}\kern-.5\wd0}}

\begin{document}

\title{Universal dynamics of a soliton after an interaction quench}

\author{Fabio Franchini}
\email{ffranchi@sissa.it}
\affiliation{INFN, Sezione di Firenze, Via G. Sansone 1, 50019 Sesto Fiorentino (FI), Italy}
\affiliation{SISSA and I.N.F.N,  Sezione di Trieste, Via Bonomea 265, 34136, Trieste, Italy}
\affiliation{Department of Physics, Massachusetts Institute of Technology, Cambridge, MA 02139, USA}

\author{Andrey Gromov}
\email{gromovand@gmail.com}
\affiliation{Department of Physics and Astronomy, Stony Brook University, Stony Brook, NY 11794, USA}

\author{Manas Kulkarni}
\email{mkulkarni@citytech.cuny.edu}
\affiliation{Department of Physics, New York City College of
Technology, The City University of New York, Brooklyn, NY 11201, USA}

\author{Andrea Trombettoni}
\email{andreatr@sissa.it}
\affiliation{CNR-IOM DEMOCRITOS Simulation Center, Via Bonomea 265, I-34136 Trieste, Italy}
\affiliation{SISSA and I.N.F.N,  Sezione di Trieste, Via Bonomea 265, 34136, Trieste, Italy}

\begin{abstract}
We propose a new type of experimentally feasible quantum quench protocol 
in which a quantum system is prepared in a coherent, localized 
excited state of a Hamiltonian.
During the evolution of this solitonic excitation, the microscopic interaction
is suddenly changed. We study the dynamics of solitons after this 
interaction quench for a wide class of systems using 
a hydrodynamic approach. We find that the post-quench dynamics is 
universal at short times, i.e. it does not depend on the microscopic details of the 
physical system.
Numerical support for these results is presented using 
generalized non-linear Schr\"odinger equation, 
relevant for the implementation of the proposed protocol with ultracold bosons, as well as for the integrable Calogero model in harmonic potential. 
Finally, it is shown that the effects of integrability breaking by a parabolic potential and by a power-law non-linearity  
do not change the universality of the short-time dynamics.
\end{abstract}


\preprint{MIT-CTP/4575}

\date{\today}

\maketitle

Recent years have witnessed a rising interest in 
non-equilibrium dynamics, which has been largely 
motivated by extraordinary progresses in the 
manipulation of many-body quantum systems 
\cite{bloch2008, greiner2002, kinoshita2006, 
joerdens2008, chen2011, trotzky2012, gring2012, 
cheneau2012, schneider2012, langen2013, smith2013, Lamacraft2011}. 
This interplay between experimental and theoretical research is 
allowing the investigation of longstanding questions, such 
as under which conditions systems driven out of equilibrium 
can show universal behaviors and whether and how a 
system reaches equilibrium and possibly thermalizes \cite{calabrese2007a, calabrese2007b, deutsch1991, srednicki1994, rigol2009, jaynes1957, rigol2007, gambassi2011, gambassi2012, sotiriadis2013, essler2012}, {  in particular in cold atom systems \cite{degrandi2010, camposvenuti2010, iyer2012, karrasch2012, dallatorre2013}.}

A particularly interesting protocol used to 
identify paradigms of out-of-equilibrium dynamics {is provided by 
the quantum quenches}, in which, typically, 
a system described by a Hamiltonian $H$ is prepared in 
its ground state and then, at a given moment of time, 
let evolved using a different Hamiltonian $H^\prime$ \cite{polkovnikov2011}.  
Although a unitary evolution cannot 
relax the system to a true equilibrium, 
mounting evidences support a picture for which, 
in most cases, the expectation value of most local operators 
tending to a stationary value in the long time limit 
can be calculated through an effective density matrix describing 
the system 
{  \cite{deutsch1991, srednicki1994, rigol2009, canovi2012, biroli2010, calabrese2012a, calabrese2012b, fagotti2013, wouters2014, pozsgay2014}}: 
that is, an arbitrary out-of-equilibrium many-body state in general 
evolves toward a state that effectively cannot be distinguished 
from a mixed state. 

In this work, inspired by these progresses and motivated 
by experimental considerations, we propose {a new protocol based 
on  a quench of the interaction on an inhomogeneous initial state localized 
in space: this protocol} has the merit of 
showing universal dynamics already at short times after the quench, {  while 
to date most of the studies on universal effects in quantum quenches 
focus on long times}. 
{We focus in particular on one-dimensional ($1d$) systems, where 
localized solitonic states are present and stable in a variety of models, 
even though we expect that our results are valid in any dimension provided 
that localized excitations exists and are stable and that the hydrodynamic 
description of the dynamics is valid. 
Specifically, we propose to prepare a $1d$}  
system in a particular 
localized, coherent excited state, one that in the hydrodynamic limit 
can be characterized as a soliton, that is, an excitation 
whose density and velocity profiles evolve in time without 
(almost) changing their shapes.
At some moment during its evolution, a control parameter 
of the microscopic Hamiltonian is changed suddenly: 
this modification of the interaction strength reflects itself, 
for instance, in the change of the scattering length and 
of the speed of sound and can be triggered experimentally 
in ultracold atom 
setups through the trapping or with an external magnetic field 
\cite{bloch2008}. After this quench, the system is let 
evolved {  according to the new Hamiltonian. 
We find that,} immediately after the quench, 
the initial profile splits into two {  ``bumps'' (over the background density): 
one moving in the same direction as the initial excitation, the other in the opposite.} 
We predict the velocities and shape {  of these chiral profiles} for short time 
after the quench and we find them to be expressible 
in a universal way. 
These findings for a global quench on a localized excitation 
have to be contrasted with the usual
quantum quench protocol discussed earlier, 
in which universality typically emerges at very large times.

{\it The hydrodynamic description: }To study the effect of an interaction quench on a localized excitation, one needs to prepare the quantum system in an excited state that would propagate for a certain time without dispersing (or dissipating). From a hydrodynamic point of view, such configuration is called a soliton. However, to date, {  the construction of a} quantum state with the degree of correlation necessary to evolve like a soliton {  has been challenging. In fact,} different attempts {  to superimpose excited states to generate stable localized excitations} have generated unstable configurations {  which in time lose coherence and fall apart \cite{kaminishi2013, delande2014}. This failure of theoretical attempts indicates that some clever insight is needed to generate a quantum state that propagates without changing its macroscopic properties, while such configurations are easily produced in the laboratories as a results of certain non-equilibrium dynamics.} Thus, for the moment we abstain from discussing the quantum nature of a solitonic excitation and we proceed from the empirical observation that such excitations are commonly observed in experiments and can nowadays be manipulated {  with remarkable precision} \cite{kevrekidis, burger99, yefsah13}.
From a mathematical point of view, solitons 
(and in particular multi-soliton solutions) are a characteristic 
feature of integrable differential equations \cite{faddeev}. 
Nonetheless, solitonic waves are commonly observed in nature, 
both in classical physics \cite{newell}, as well as 
a collective behavior of a quantum many-body systems \cite{kevrekidis}, 
proving that, in certain regimes, one can neglect all 
other contributions and describe the system accurately 
by an integrable equation, 
thus justifying the existence of very long-lived excitations.


{We will focus on systems whose quantum dynamics can be described 
by a hydrodynamic approach: this approach is successfully used for a wide 
class of systems \cite{forster} 
including in particular trapped ultracold atoms 
\cite{stringari}. The hydrodynamics of a quantum system may be derived, for instance, from an effective single-particle mean-field equation, provided  
a functional energy is guessed or derived (this point of view is typical of density functional theory 
approaches \cite{ullrich}): 
a major example is given by weakly interacting 
Bose gases where the Non-Linear Schr\"odinger equation (NLSE) 
provides equilibrium 
\cite{lieb} and dynamical \cite{erdos2007} properties of the corresponding quantum system.
 
For quantum systems for (and the regimes in) 
which the hydrodynamic approach is valid, one can write 
as usual \cite{forster} the} continuity and Euler equations 
for the density and velocity fields
\bea
   \label{cont}
   && \dot \rho + \p (\rho v) = 0; \: \: \: \: \dot v + \p {\cal A}=0\quad \\
   \label{Euler} 
   && {\cal A} \equiv \frac{v^2}{2} + \omega(\rho) - A^\prime(\rho)(\p\sqrt{\rho})^2 - A(\rho)       
   \frac{\p^2\sqrt{\rho}}{\sqrt{\rho}},
\eea
where $\omega (\rho)$ is the specific enthalpy and $A(\rho)$ 
is related to the quantum pressure{: both $A$ and $\omega$ are 
model-dependent.}


Linearization of the hydrodynamic equation around a constant background 
leads to sound waves (the Bogolioubov modes).
As {derived in \cite{novikov1984,kulkarni2012}, in $1d$} 
for configurations with small, smooth perturbations over 
the background $\rho_0$ {it is} possible to retain the 
leading non-linear and dispersive terms, which, 
under the condition of locality of the interactions, 
have the universal form of the KdV equation:
\be
   \label{2KdV}
   \dot{u}_\pm \mp \p_x \left[ cu_\pm + {\zeta \over 2} u_\pm^2 - \alpha \p_x^2 u_\pm \right] = 0 \; ,
\ee
where $c \equiv \sqrt{ \rho_0 \omega'_0}$ is the sound velocity 
($\omega'_0 \equiv \partial_\rho  \omega|_{\rho_0}$) and the nonlinear 
and dispersive coefficients are given by
\be
   \zeta \equiv  \frac{c}{\rho_0} + \frac{\p c}{\p \rho_0} \; , \qquad
   \alpha \equiv \frac{A (\rho_0) }{4c} \; . 
   \label{zetaalphadef} 
\ee
Here, a formal (small) expansion parameter $\epsilon$ was introduced, 
so that the fields could be expanded around the static background as
\bea
   \rho (x,t) &=& \rho_0 + \epsilon \: \rho^{(1)} (x,t) + \epsilon^2 \: \rho^{(2)} (x,t) + \ldots \; , \\
   v(x,t) &=& \epsilon \: v^{(1)}  (x,t) + \epsilon^2  \: v^{(2)}  (x,t) +\ldots  \; .
\eea
The first order terms evolve according to the KdV:
\be
   u_\pm = \rho^{(1)} = {c \over \omega'_0} \: v^{(1)} \; .
\ee
Since KdV is a chiral equation, the $\pm$ refers to the 
two chiral components of the fields. We neglected the 
interaction between the left and right moving sectors, 
since locality implies that this effect is relevant 
only when the two are overlapping, but this happens 
for short times, as they pass through each other 
with a relative velocity of approximately $2c$ \cite{novikov1984}.

{\it The interaction quench: }
Let us denote with $g$ some microscopical 
parameter capturing the strength of the interaction 
in the quantum Hamiltonian. Suddenly, during the 
evolution of the soliton, we change the interparticle 
coupling to $g'$, which reflects itself in the change 
of the effective parameters $\omega$ and $A$ in 
(\ref{cont}, \ref{Euler}) and hence of $c$, $\zeta$, and 
$\alpha$ in (\ref{2KdV}). We 
denote with a prime all the post-quench parameters, 
calculated using $g'$. As a consequence of this global quantum quench, 
the initial profile 
is not anymore a soliton 
of the new system and hence will not be stable under the new evolution.

\begin{figure}
\begin{center}
\includegraphics[width=\columnwidth]{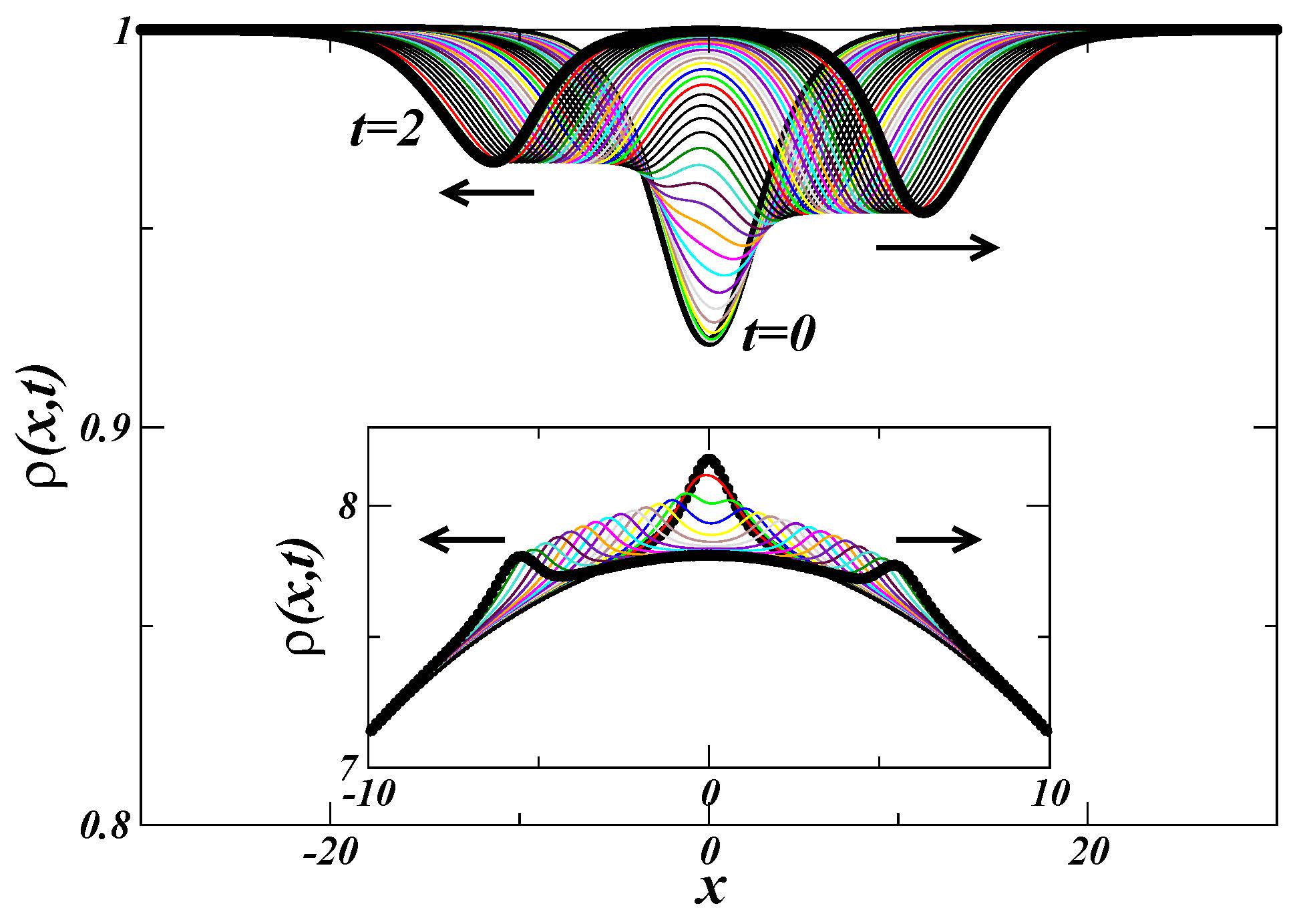}
\caption{Plot of the NLSE density $|\psi(x,t)|^2$ for different times 
$t=0,0.05, 0.1, \cdots, 2,$ with $c'/c=40$ and initial velocity $V=0.96c$ 
($\hbar=m=\rho_o=1$, where $\rho_o$ is the $x \to \pm \infty$ density). 
The interaction is suddenly changed at $t_Q=0.1$. Inset: plot 
of the density for the Calogero model at times $t=0, \Delta t, 2 \Delta t, 
\cdots, 15 \Delta t$ with $\Delta t=0.157$ and $c'/c=10$ (moreover,  
$N=301$, $c=0.23$, $V=1.002c$, $t_Q=0$). 
In both cases the first and last densities are 
evidenced by a thick line and only the central regions of the $x$-axis have 
been showed.}  
\label{fig:evolution}
\end{center}
\end{figure}

The profile experiences the quench as an external perturbation, 
to which it reacts by splitting into a transmitted and reflected component, 
exactly as it would happen to a linear wave, if the sound velocity 
was suddenly changed. A typical non-linear quench dynamics of 
this type for the dark soliton of the NLSE is shown 
in Fig. \ref{fig:evolution}: 
one profile moves forward in the same direction as the initial soliton 
(the transmitted one), while the other moves in the opposite direction 
(the reflected one). 

Hence, we make the ansatz that immediately after the quench we have
\be
   u (x,t) = u^r (x - V_r t) + u^t (x - V_t t) \; ,
   \label{ansatz}
\ee
assuming that for short times after the quench 
the two chiral profiles evolve without changing 
their shapes significantly. If the initial soliton 
has $V > 0$, the reflected velocity $V_r < 0 $ and 
the transmitted one is $V_t  > 0$. 
Imposing continuity of the solution and the conservation 
of momentum at $t=0$ constraint the post-quench configuration to be 
\be
   u(x,t) = R  \:  s (x - V_r t) + T \: s (x - V_t t) \; ,
   \label{usol}
\ee
that is, we find that the two chiral profiles maintain the same 
functional shape as the original soliton, but with 
their heights reduced by the reflection and transmission coefficients, 
which are found to be
\be
   R (V, V_r, V_t) = {V_t - V \over V_t - V_r}  \; , \quad 
   T (V, V_r, V_t) = {V - V_r \over V_t - V_r} \; .
   \label{RTdef}
\ee
We remark that, so far, we only applied kinematic considerations 
(we made no use of the KdV), {which alone are not sufficient to determine} the velocities $V_{r,t}$. 

In the corresponding linear problem, 
all waves would move at the speed of sound, 
that is $V=c$ and $V_t = - V_r = c'$. 
Hence, sound waves would be scattered by an interaction quench with 
\be
   R_{\rm linear} = {1 \over 2} \left[ 1 - {c \over c'} \right] \; , 
   \qquad    T_{\rm linear} = {1 \over 2} \left[ 1 + {c \over c'} \right] \; .
   \label{RTlinear}
\ee
Naturally, for a small quench ($c' \simeq c$) the reflected wave has 
vanishing amplitude, while for large quenches ($c' \gg c$) both waves are 
comparable.

The non-linear terms in (\ref{2KdV}) make the different profiles 
move at different velocities. The stability of 
the initial soliton is given by a balancing of opposite effects, 
coming from the non-linear and dispersive terms in (\ref{2KdV}). 
It can be shown that the average velocity of the solitonic solution is that of its barycenter, 
where the dispersion is perfectly balanced and $V$ is determined only 
by the non-linear contribution. Using this observation, we can estimate the 
velocities of the two chiral profiles to be \cite{suppl}
\bea
   V_r & = & - \left[ c - \eta \:  R \: (c - V) \right] {c' \over c} \; ,
   \label{dVcmR} \\
   V_t & =& \quad \left[  c  - \eta \: T \: (c - V)  \right] {c' \over c} \; , 
   \label{dVcmT}
\eea
where we introduced the universal parameter 
\be
   \eta \equiv {c \over c'} \: {\zeta' \over \zeta} 
    = {1 + {\rho_0 \over c'} \: {\partial c' \over \partial \rho_0} \over
    1 + {\rho_0 \over c} \: {\partial c \over \partial \rho_0} } \; ,
   \label{etadef}
\ee
and where the $T$ and $R$ are found consistently to be 
\bea
   R & =&  {1 \over 2} \left[ 1 - {c \over c'} \: { V \over \eta \:  V  + (1 - \eta) \: c  } \right] \; ,
   \label{Rest} \\
   T & = & {1 \over 2} \left[ 1 + {c \over c'} \: { V \over \eta \:  V  + (1 - \eta) \: c  } \right] \; .
   \label{Test}
\eea
We note that these expressions are completely universal and 
depend on the microscopics of the system {only through the parameter 
$\eta$, which in turn depends on $A$ and $\omega$ 
($\eta \simeq 1$ for the $1d$ NLSE)}. 
In particular, everything can be made dimensionless by 
measuring velocities in unit of the sound velocity 
(for instance, before the quench). 
Experimentally, the velocities and heights of the 
two profiles can be measured and checked against our predictions 
through a time-of-flight experiment, once $\eta$ 
is measured using (\ref{etadef}).

If we assume that 
$\omega (\rho)$ in (\ref{Euler}) is a simple monomial 
of the density ($\omega (\rho) \propto \rho^{\kappa-1}$), 
then $\eta=1$ and the formulae simplify to
\bea
   R  =  {1 \over 2} \left[ 1 - {c \over c'} \right] \; , & \quad &
   T  =  {1 \over 2} \left[ 1 + {c \over c'} \right] \; ,
   \label{RT} \\
   V_r =  - \left( T \: c +  R \: V  \right) {c' \over c}  \; , & \quad &
   V_t = \left( R \: c  +  T \:  V \right) {c' \over c}  \; .
   \label{dVcm}
\eea
We note that these reflection and transmission coefficients
are the same as of the linear process (\ref{RTlinear}), 
although accompanied by non trivial velocities for the profiles.

{\it $1d$ ultracold bosons: }
{Localized excitations are easy to prepare in a cold atom experiments: 
using a phase mask it is possible to create long-lived 
excitations that can 
travel through the system without significantly 
changing their shape. The initial momentum can be  given 
by suitably varying the trap potential, as it has been done to induce 
and study the dynamics of dark solitons in ultracold bosonic \cite{burger99} 
or fermionic \cite{yefsah13} condensates. Moreover, the interaction 
between bosons can be  varied by changing the scattering length 
\cite{stringari}: our 
interaction quench can be accomplished by a rapid change of the  magnetic field, 
easily feasible in present-day experiments. Moreover, the splitting 
of the solitons can be straightforwardly detected by interference patterns measurements.}

{We write the $1d$ NLSE in the generalized form
\be
   i \hbar \partial_t {  \psi} = \left\{ - {\hbar^2 \over 2 m} \partial_{xx} 
   + f(\rho) \right\} \psi \; ,
   \label{NLS}
\ee 
where  $\rho (x,t) = \left| \psi (x,t) \right|^2$. The choice $f(\rho)=g\rho$ 
corresponds to the usual NLSE describing bosons with contact interaction 
in the weakly interacting 
limit. The generalized NLSE  (GNLSE) reduces to (\ref{cont},\ref{Euler}) 
with the ansatz ${  \psi} = \sqrt{\rho} e^{i {m \over \hbar} \int^x v(x') d x'}$, 
with $\omega(\rho) = \frac{f(\rho)}{m}$ and $A = \frac{\hbar^2}{2m^2}$ 
(of course $\omega(\rho) = \frac{g\rho}{m}$ for the  usual  NLSE). 
Several $1d$ classical mean-field equations have been proposed and used 
to reproduce the quantum dynamics of the $1d$ Bose gas for not so small 
interactions \cite{kolomeisky00,salasnich2002,kim03,choi2014} and agreement 
between NLSE results, dynamics of the quantum model 
and experimental findings is found, and expected  to be more robust 
at {{\em short} times, that is, the regime we are considering}. {  In particular short times have to be considered 
the ones much smaller than $\hbar/\Delta E$, where $\Delta E$ is the energy difference between the expectation 
values of the Hamiltonian post- and pre-quench, even though numerical simulations show that the results are stable also for longer times.} 

The validity of 
the hydrodynamic approach for the $1d$ Bose gas has been shown 
in \cite{menotti2002,choi2014}, while the hydrodynamic equations 
(\ref{cont},\ref{Euler}) have been derived for $1d$ ultracold systems 
in \cite{kulkarni2012}. We observe that in our approach the NLSE and 
the GNLSE are treated on equal footing since 
the analytical results of the GNLSE (\ref{NLS}) are obtained using 
the correct $\omega(\rho)$ in the hydrodynamic equations.}

\begin{figure}
\begin{center}
\includegraphics[width=\columnwidth]{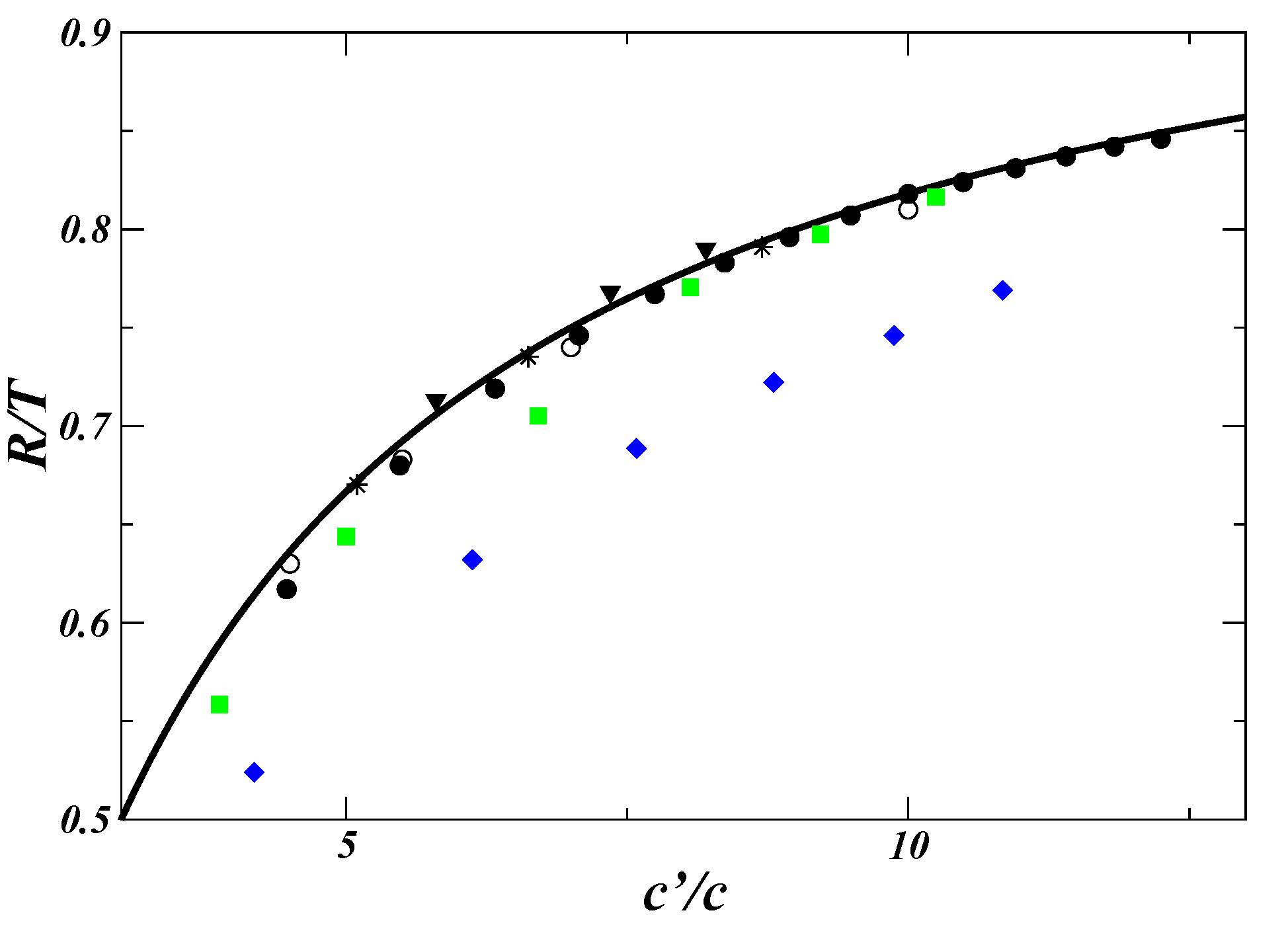}
\caption{Height ratio of the chiral profiles vs. $c'/c$. The black solid lines is our 
analytical prediction (\ref{RT}), plotted 
together with numerical results: the initial velocity 
of the grey soliton is $V=0.96c$ 
(black circles), $V=0.9s$ (green squares). 
Asterisks refers to the case with a parabolic potential ($\omega=0.01$ and 
$V=0.96c$), while triangles refers the the NLSE with a power-law 
non-linearity ($\alpha=2.2$ and 
$V=0.96c$) - in all cases $t_Q=0.02$. Open circles refer 
to the  corresponding data for the Calogero model and the same parameters 
of Fig. \ref{fig:evolution}. The outliers blue diamonds correspond to NLSE with a large initial soliton ($V=0.5c$), where our approximations are suppose to be less accurate.} 
\label{fig:heights}
\end{center}
\end{figure}
\begin{figure}
\begin{center}
\includegraphics[width=\columnwidth]{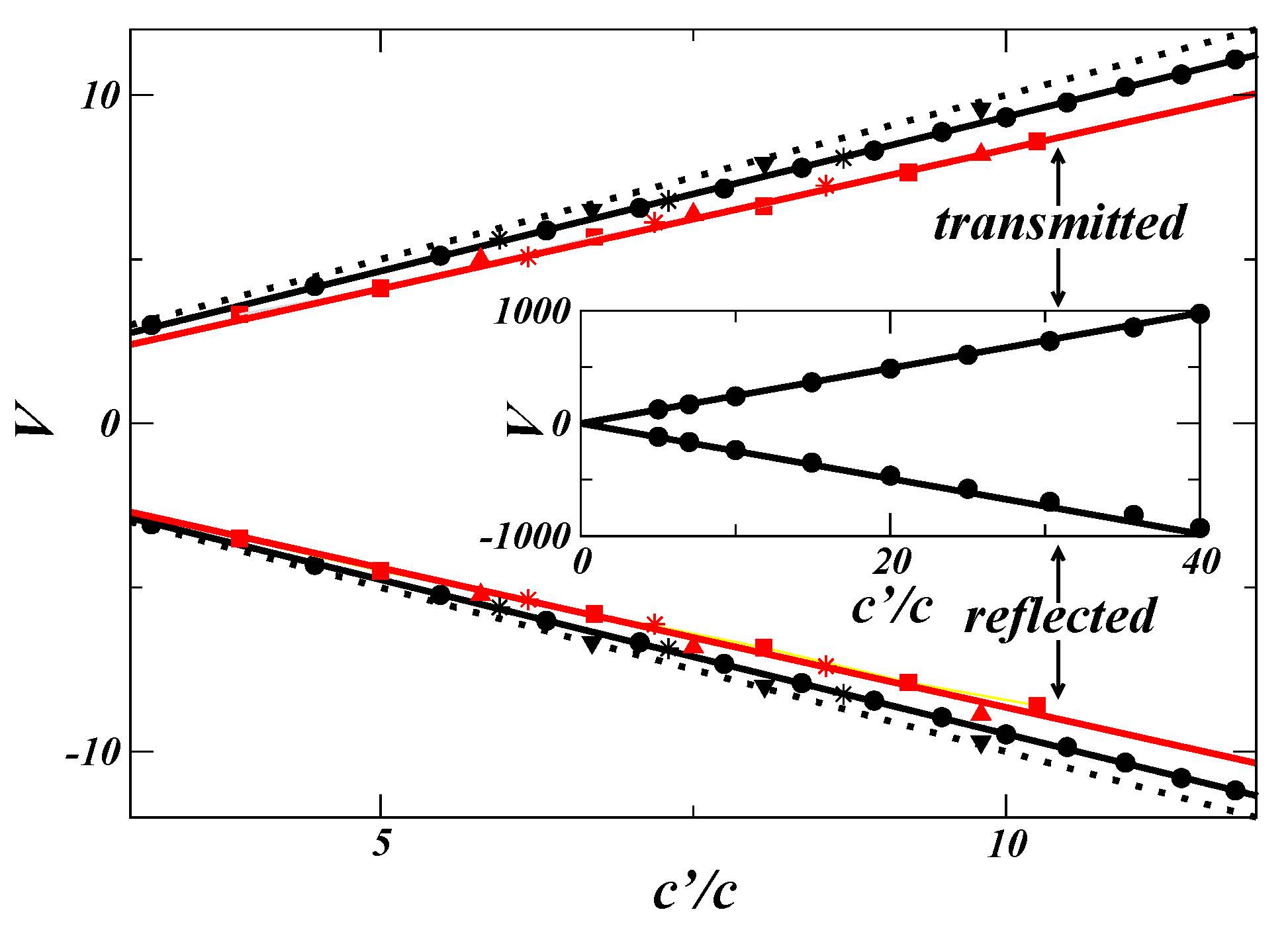}
\caption{Reflected (bottom) and transmitted (top) 
peak velocities vs $c'/c$. Black and 
red lines and points refer respectively to initial soliton velocities 
$V=0.96c$ and $V=0.9c$: lines are our analytical prediction, while red circles 
and black squares refers to NLSE numerical results. For comparison we also 
plot $\pm c'/c$ (dashed lines). Asterisks and triangle refer respectively to NLSE with a parabolic potential and to the NLSE with a power-law  (same parameters as in Fig. \ref{fig:heights}).
Inset: peak velocities reflected (top) and 
transmitted for the Calogero model (same parameters as of Fig. \ref{fig:evolution}).}
\label{fig:velocities}
\end{center}
\end{figure}

{Therefore, without loss of generality, in the following we present our results for the dark soliton of the 
NLSE \cite{novikov1984} subjected to the interaction quench.} 
In Figs. \ref{fig:heights}--\ref{fig:velocities} we plot the results of our 
numerical 
simulations vs. the analytical predictions discussed so far, 
finding a remarkable agreement. 
As the reflected and transmitted profiles are created superimposed 
at the time of the quench, they need to evolve for a certain time 
before they can be distinguished. In this time, their shapes 
evolve (and they also interact with one-another) and this introduces 
additional noise in the simulations. Moreover, in this way we measure the velocity of the peaks of the profile, which could differ from the center-of-mass velocity due to dispersion, since these profiles are not solitons \cite{foot1}.
Nonetheless, a remarkable good agreement is found, which gets better for larger interaction quenches, since the dispersion effects we neglect are expected to become smaller. Furthermore, as expected 
for darker grey solitons (i.e., for larger solitons) the agreement becomes 
worse. 


In Fig. \ref{fig:heights} we plot the measured ratio between 
the heights of the two chiral profiles vs. the ratio $R / T$ from (\ref{RT}) 
for different quenches and with different velocities of the initial soliton, while in Fig. \ref{fig:velocities} 
we compare the measured peak velocities with (\ref{dVcm}).
In all the numerical points the values are computed at the mid-time between 
the instant in which there are $2$ bumps and the one in which a 
third bump emerges \cite{suppl}.  
In addition to the NLSE, to test the universality of our results, we also computed the same 
quantities using a parabolic confinement (i.e., adding $V\psi$ to the NLSE 
with $V=(1/2) \omega^2 x^2$) or a power-law NLSE (i.e., having a non-linear term 
of the form $|\psi|^\alpha\psi$): both are non-integrable deformations 
of the NLSE and support the universality of our results \cite{foot3}.

{\it Calogero model: } 
{As a further check on our ansatz (\ref{ansatz}), 
we compare our predictions against 
a microscopical evolution. The rationale 
for such check is that, even if we know that for a (classical 
or quantum) system the hydrodynamic 
description is  valid, 
our ansatz (\ref{ansatz}) may be wrong in presence of general non-local 
interactions.} For this experiment, 
we use the classical Calogero model, given by the Hamiltonian
\be
    H = {1 \over 2 m} \sum_{j=1}^N p_j^2 + {\hbar^2 \over 2 m} 
    \sum_{j \ne k} {\lambda^2 \over (x_j - x_k)^2} + {\omega^2 \over 2} \sum_{j=1}^N x_j^2 \; ,
\ee
with a dimensionless coupling constant $\lambda$. This is a classical model 
which is integrable even in the presence of an external parabolic potential. 
Its hydrodynamic description is thus not the KdV, but a different 
integrable equation, in the family of the Benjamin-Ono (BO) {equation}
{\cite{abanov2009}}. 
It differs from the KdV by its dispersive term and by the fact that its 
solitons have longer (power-law) tails. Another difference is that this 
system supports supersonic bright solitons, instead of the subsonic dark 
ones of local models \cite{AGK}. Nonetheless, it has $\eta =1$ and we 
find that the center-of-mass 
velocities and transmission/reflection coefficients have the same universal 
expression (\ref{RT},\ref{dVcm}). 
In our numerical simulation, we simulated the Newtonian evolution of $N=301$ 
particles
which initially evolve as a soliton of the pre-quench system and we follow 
their dynamics after the quench \cite{foot4}. 
We notice the correct splitting of 
the density field into the two chiral profile and we see from 
Figs. \ref{fig:heights}--\ref{fig:velocities} 
that this microscopic dynamics agrees 
remarkably well with our analytical predictions. {We stress that, although
in this case the numerical experiment simulates a classical, Newtonian particle 
dynamics, it still shows a clear chiral splitting of the profiles due to the quench.}

{\it Conclusions: }
We have proposed a novel quantum quench protocol, 
where we initially prepare the system in a localized excitation and 
evolve it after changing the interaction strength {and we studied 
this interaction quench of quantum systems using a hydrodynamic 
approach.} 
We found that the dynamics immediately after the quench is universal, 
i.e. it does not depend on the details of interaction, 
but only on the strength of the quench, measured from macroscopic parameters, 
such as the speed of sound, see (\ref{dVcmR}, \ref{dVcmT}, \ref{Rest}, \ref{Test}). 
We checked our analytical predictions against the numerical simulation 
of the $1D$ NLSE also in presence of a parabolic potential and with 
power-law non-linearity) 
and through the Newtonian evolution of the Calogero model 
in harmonic potential finding a remarkable agreement. 
We finally remark that the universal nature of our results implies 
that this protocol can also have {ubiquitous applications in quantum optics, 
nonlinear waveguides and in other non-linear systems}.\\

{\it Note Added:} During the final stages of this work, 
we were made aware of a parallel development by 
O. Gamayun {\em et al.} \cite{gamayun2014}, 
concerning the long time behavior of the NLSE after the quench protocol 
we discussed, where the integrability of 
the hydrodynamics leads to remarkable effects. 
After the submission 
of our paper {  on arXiv, a  paper by A. Chiocchetta {\em et al.} appeared \cite{chiocchetta2014} 
discussing the universality of the scaling at short times 
of a quantum system after a quench.}

{\it Acknowledgements:}
We acknowledge discussions with S. Sotiriadis{, D. Schneble and A. Polychronakos}.
FF was supported by a Marie Curie International Outgoing 
Fellowship within the 7th European Community Framework Programme 
(FP7/2007-2013) under the grant PIOF-PHY-276093. MK gratefully acknowledges support from the Professional Staff Congress – City University of New York award \# 68193-00 46.


\begin{thebibliography}{99}
\bibitem{foot1}
   It is possible to include this effect by accounting for dispersion and the estimate for the peak velocity are also expressible in universal terms.
\bibitem{foot3}
   The data for the power-law NLSE are generated starting from the integrable NLSE soliton. This approximation and the effect of dispersion are probably accountable for some deviation 
from the expected behavior in Fig. \ref{fig:velocities}.
\end{thebibliography}
\end{document}